\documentclass[twocolumn,showpacs,preprintnumbers,amsmath,amssymb]{revtex4}
\usepackage[dvips]{graphicx}
\usepackage{dcolumn}
\usepackage{bm}    

\begin{document}
\title{Synchronization of coupled generators of quasiperiodic oscillations}
\author{Nataliya~V.~Stankevich$^{1,*}$, Alexander~P.~Kuznetsov$^2$}
\affiliation{$^{1}$ Yuri Gagarin State Technical University of Saratov, \\
Politechnicheskaya 77, Saratov, 410077, Russian Federation\\
$^{2}$Kotel'nikov's Institute of Radio-Engineering and Electronics of RAS, Saratov Branch,\\
Zelenaya 38, Saratov, 410019, Russian Federation}

\date{\today}
\begin{abstract}
The dynamics of two coupled generators of quasiperiodic
oscillations is studied. The opportunity of complete and phase
synchronization of generators in the regime of quasiperiodic
oscillations is obtained. The features of structure of parameter
plane is researched using charts of dynamical regimes and charts
of Lyapunov exponents, in which typical structures as resonance
Arnold web were revealed. The possible quasiperiodic bifurctions
in the system are discussed.
$^{*}$Corresponding author. Tel.:+7 9033290994; fax: +7 8452998810

E-mail address: stankevichnv@mail.ru (N.V.Stankevich)
\end{abstract}


\maketitle

\section{Introduction}
Quasiperiodic oscillations are well-known for the applications in different science and technic \cite{r1}-\cite{r8}. Therefor the problem of synchronization of this type of oscillations is a highly examinate research question in nonlinear dynamics. The interesting problem here is to analyze the task of synchronization of quasiperiodic oscillations, which is less studied than chaotic and periodic oscillations. However, the full theory of synchronization of quasiperiodic oscillations allows us to consider the broad class of possible problems.

In particular, it can be task about dynamics of two coupled autonomous system with quasiperiodic oscillations. Nowadays, there no so many examples of quasiperiodic generators, which is suitable for application nonlinear dynamics methods. As the first it could be mentioned one of the modification of the Chua circuit presented in \cite{r8}. However, it is described by piecewise-linear characteristic and, therefor, it is not well studied in the context of the problem of the synchronization of quasiperiodic oscillations \cite{r8}.

 Recently was proposed a modification of Anishchenko-Astakhov generator with autonomous quasiperiodic dynamics \cite{r10}-\cite{r12}, as a result of adding a new variable to a classic generator of Anishchenko-Astakhov. In this case the dimensional of phase space will be equal four. For autonomous system was discovered such a significant result as a doubling of invariant tori \cite{r10}. Other important founding were shown for the system of two coupled generators \cite{r11}-\cite{r12}. For example, was revealed the picture of occurrence of the resonance of invariant tori on the surface of high-dimensional torus. This picture corresponds to cascade of sequence saddle-node bifurcations of invariant tori. With help of the spectrum of the Lyapunov exponent was illustrated how the increasing of frequency detuning in the system lead to the sequence birth of many-frequency torus from two-frequency. This effect was discovered in electronic experiment. However, the cases of studied were limited by the weak coupling and small frequency detuning. Moreover, in this research only one-parametric analysis is applied. At the same time, it is well known, how important is to consider the case of two-parametric picture, when one analyzes coupled generators.

 In the current work we provide two-parametric analysis of the synchronization of quasiperiodic generators in the wide range of changing of the controlling parameters. We work with simpler model of generator of quasiperiodic oscillations with three-dimensional phase space \cite{r14}, \cite{rn15}, \cite{rn16}.

\section{Coupled generators of quasiperiodic oscillations}

Now we describe a system, which was presented in \cite{r14} and which we will use as a generator of quasiperiodic oscillations. This system is the "hybrid" of a self-generator with a hard excitation and a relaxation generator and has a following view::
\begin{equation}
\label{1GQP}
  \begin{array}{l l}
    \ddot {x}-(\lambda+z+x^{2}-\beta x^{4})\dot{x}+\omega_{0}^{2}x=0,\\
    \dot {z}=\mu-x^{2},
  \end{array}
\end{equation}
where $\omega_{0}$ ($\omega_{0}>0$) is main frequency of the generator, $\lambda$ ($\lambda\geq0$) is the parameter of excitation (negative friction), $\beta$ ($\beta>0$) is the response for the saturation of oscillations at large amplitudes. The parameter $\lambda$ enters to the equation as well as $z$, which characterizes the state of a charging relaxation element, and its evolution in time is controlled by the second equation. The system (\ref{1GQP}) has two independent time scales that allows for a two-frequency quasiperiodic dynamics. The first time-scale is the period of oscillations of the self-generator $T=\frac{2\pi}{\omega_{0}}$ and the second one is the characteristic recovery time of the state storage element $\tau=\mu^{-1}$.

\begin{figure}[h]
\begin{center}
\includegraphics[scale=0.5]{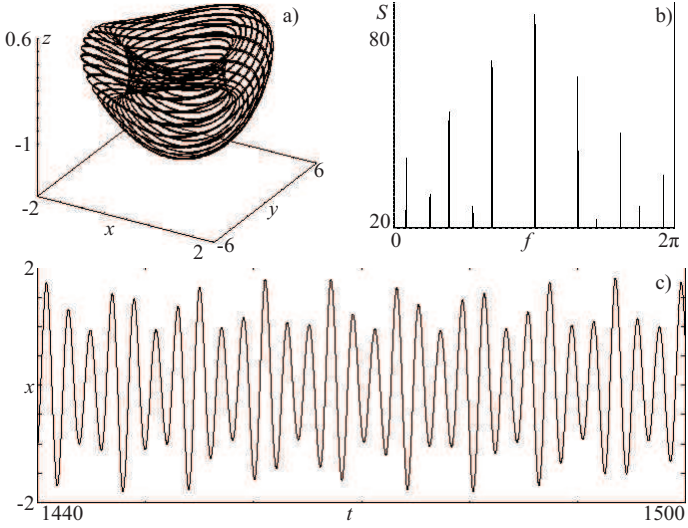}
\end{center}
\caption{Phase portrait in the form of invariant torus a),
Fourier spectrum b) and time realization с) of the system (\ref{1GQP}) in quasiperiodic regime; $\lambda=0$, $\mu=0.9$, $\omega_{0}=\pi$.} \label{Fig.1}
\end{figure}
In Fig.~\ref{Fig.1} are shown the example of phase portrait of the system (\ref{1GQP}) in the form of invariant torus, quasiperiodic time realization $x(t)$ and Fourier spectrum of quasiperiodc regime.

Now let consider a  system of two coupled generators of this kind:
\begin{equation}
\label{2GQP}
  \begin{array}{l l l l}
    \ddot {x}_{1}-(\lambda+z_{1}+x_{1}^{2}-\beta x_{1}^{4})\dot{x}_{1}+\omega_{01}^{2}x_{1}+M_{C}(\dot{x}_{1}-\dot{x}_{2})=0,\\
    \dot {z}_{1}=\mu_{1}-x_{1}^{2},\\
    \ddot {x}_{2}-(\lambda+z_{2}+x_{2}^{2}-\beta x_{2}^{4})\dot{x}_{2}+\omega_{02}^{2}x_{2}+M_{C}(\dot{x}_{2}-\dot{x}_{1})=0,\\
    \dot {z}_{2}=\mu_{2}-x_{2}^{2}.
  \end{array}
\end{equation}

Here $x_{1}, z_{1}$ are variables, which characterize the first generator, and $x_{2}, z_{2}$ are variables of the second generator, $M_{C}$ is a coefficient of dissipative coupling.

The system (\ref{2GQP}) is characterized by four independent frequencies, which are determined by a four controlling parameters: the base frequencies of the generators $\omega_{01}, \omega_{02}$ and frequencies of relaxation oscillations $\mu_{1}, \mu_{2}$. We will consider two cases of changing of this parameters.

When we consider the problem of the synchronization of quasiperiodic oscillations, it is important to be sure that the autonomous quasiperiodic regime is stable to changing the controlling parameters. We can use that property of autonomous system (\ref{1GQP}), that at $\omega_{0}>3$ the resonance tongues became very narrow, and quasiperiodic regimes dominate \cite{r14}. Therefor we fix $\omega_{01}=\omega_{0}$, $\omega_{02}=\omega_{0}+\Delta$, where $\Delta$ represents controlling frequency detuning. Other parameters we consider to be equal: $\omega_{0}=\pi$, $\mu_{1}=\mu_{2}=0.9$. For this parameters the first generator in autonomous regime demonstrates quasiperiodic oscillations. Also the second subsystem, mainly, behaves in the regime of autonomous quasiperiodicity with the increasing of frequency detuning $\Delta$.

\begin{figure}[h]
\begin{center}
\includegraphics[scale=0.5]{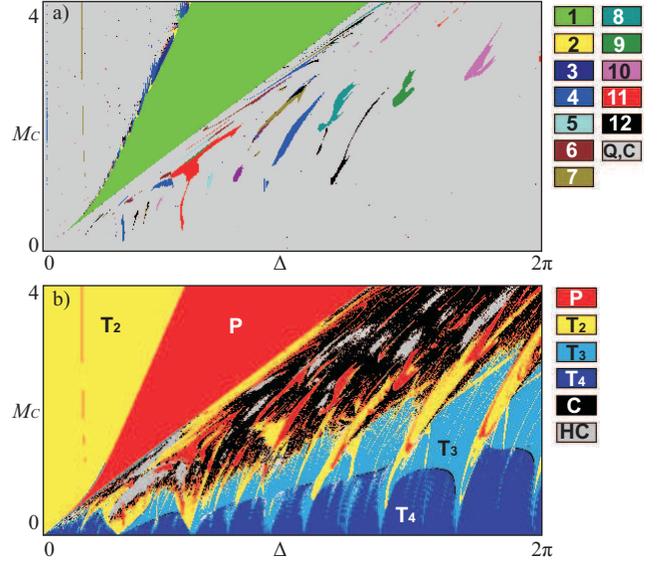}
\end{center}
\caption{a) Chart of dynamical regimes, b) chart of Lyapunov exponents of coupled generators (\ref{2GQP}); $\lambda=0$,
$\omega_{0}=\pi$, $\mu_{1}=\mu_{2}=0.9$.}
\label{Fig.2}
\end{figure}

In Fig.~\ref{Fig.2}~a) the chart of dynamical regimes for the system of coupled generators (\ref{2GQP}) on the parameter plane the frequency detuning vs coefficient of coupling ($\Delta$, $M_{C}$) in the wide range of changing parameters is presented. The regular regimes of different periods are denoted by different colors, the color palette is interpreted in the right part of figure. Non-periodic (chaotic and quasiperiodic) regimes are shown by gray color \footnote{The period of oscillations determine as/like the period of cycle in the Poicar\'{e} section by hypersurface $\dot{x}_{1}=0$}.

For more detailed description of the dynamics of the system (\ref{2GQP}) also the chart of Lyapunov exponents was constructed. We show it on Fig.~\ref{Fig.2}~b). For this chart the color palette was chosen in corresponding with signature of the spectrum of Lyapunov exponents, so that the following regimes can be visualized:\\
periodic regime $P$, $\Lambda_{1}=0>\Lambda_{2}>\Lambda_{3}>\Lambda_{4}>\Lambda_{5}>\Lambda_{6}$;\\
two-frequency quasiperiodic regime $T_{2}$, $\Lambda_{1}=\Lambda_{2}=0>\Lambda_{3}>\Lambda_{4}>\Lambda_{5}>\Lambda_{6}$;\\
three-frequency quasiperiodic regime $T_{3}$, $\Lambda_{1}=\Lambda_{2}=\Lambda_{3}=0>\Lambda_{4}>\Lambda_{5}>\Lambda_{6}$;\\
four-frequency qusiperiodic regime $T_{4}$, $\Lambda_{1}=\Lambda_{2}=\Lambda_{3}=\Lambda_{4}=0>\Lambda_{5}>\Lambda_{6}$;\\
chaotic regime $C$, $\Lambda_{1}>\Lambda_{2}=0>\Lambda_{3}>\Lambda_{4}>\Lambda_{5}>\Lambda_{6}$;\\
hyperchaotic regime $HC$, $\Lambda_{1}>\Lambda_{2}>\Lambda_{3}=0>\Lambda_{4}>\Lambda_{5}>\Lambda_{6}$.

For beginning, we discuss the largest area of two-frequency regimes, denoted by symbol $T_{2}$ on Fig.~\ref{Fig.2}~b). This area has form of Arnold tongue with bottom in a point $\Delta=0$,
$M_{C}=0$. Inside this area exists the regime of \emph{phase synchronization of quasiperiodic oscillations}. For more visible results we provide here detailed analysis.

\begin{figure}[h]
\begin{center}
\includegraphics[scale=0.5]{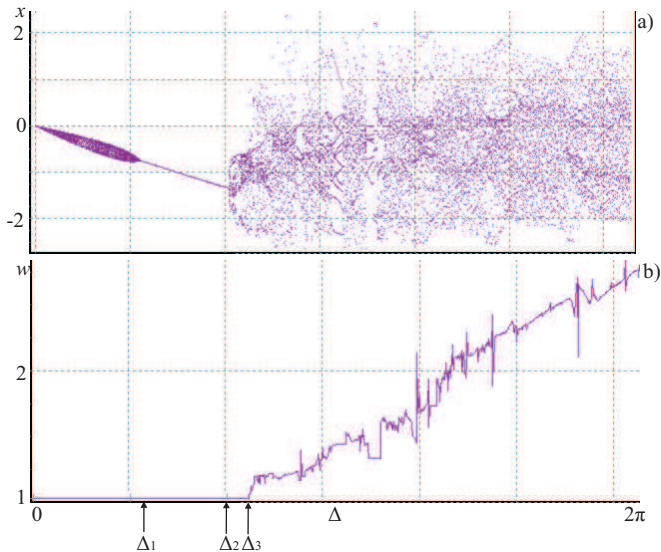}
\end{center}
\caption{Bifurcation tree and the dependence of winding number on the frequency detuning for $\lambda=0$, $\omega_{0}=\pi$, $\mu_{1}=\mu_{2}=0.9$,
$M_{C}=2$.} \label{Fig.3}
\end{figure}

On Fig.~\ref{Fig.3} is presented a bifurcation tree. It is the dependence of the variables $x_{1}$ on the frequency detuning $\Delta$ in the Poincar\'{e} section by hypersurface $\dot{x}_{1}=0$ at the fixing coupling $M_{C}=2$. The bottom figure shows the dependence of winding number $w$ on the parameter  $\Delta$, which was calculated by the following equation:
\begin{equation}
\label{RN}
  w=\lim_{t\rightarrow\infty}\frac{N_{t}^{x_{1}}}{N_{t}^{x_{2}}},
\end{equation}
where $N_{t}^{x_{1}}$ is amount of intersections of hypersurface $x_{1}=0$ by phase trajectory, $N_{t}^{x_{2}}$ is amount of intersections of hypersurface $x_{2}=0$. As it can be seen from the bifurcation diagram, in the system non-regular, quasiperiodic regime is observed at $\Delta<\Delta_{1}$ . On the dependence of the winding number this interval corresponding constant winding number $w=1$. Thus, generators are mutually locking, but their oscillations are two-frequency quasiperiodic.

\begin{figure}[h]
\begin{center}
\includegraphics[scale=0.5]{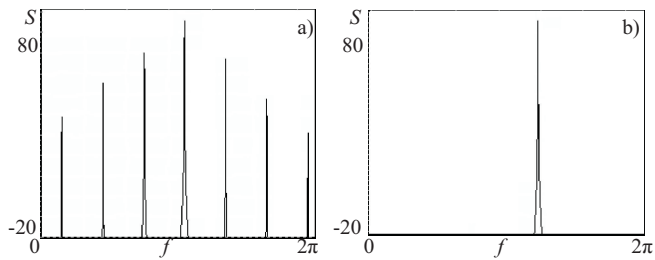}
\end{center}
\caption{Fourier spectrums for system (\ref{2GQP}) at $\lambda=0$,
$\omega_{0}=\pi$, $\mu_{1}=\mu_{2}=0.9$, $M_{C}=2$; a)
$\Delta=0.5$ и b) $\Delta=1.5$.} \label{Fig.4}
\end{figure}

If we increase the frequency detuning $\Delta_{1}<\Delta<\Delta_{2}$ then periodic regime occurs. In the chart of dynamical regimes (Fig.~\ref{Fig.2}~a)) it corresponds to the area of period-1 marked by green color. Thus, when we go outside from main tongue of two-frequency quasiperiodicity, at the large coupling, \emph{the complete synchronization} of generators occurs, and it responds the simplest periodic regimes. We can say, that at enough large coupling and enough large frequency detuning one can observe suppression of quasiperiodic oscillations due to generators interaction. Fig.~\ref{Fig.4} demonstrates the evolution of the Fourier spectrum at the transition from quasiperiodic phase synchronization to the complete synchronization with locking of all frequency components. The frequency of the occurring main component can be estimated as:
\begin{equation}
\label{eq_Freq}
  \omega_{0}^{ср}=\frac{\omega_{01}+\omega_{02}}{2}=\omega_{0}+\frac{\Delta}{2}.
\end{equation}

Further, in a few interval in the range $\Delta_{2}<\Delta<\Delta_{3}$, how it can see in Fig.~\ref{Fig.4}~a), there is phase synchronization, but the regime again becomes non-periodic (two-frequency).

\begin{figure}[h]
\begin{center}
\includegraphics[scale=0.6]{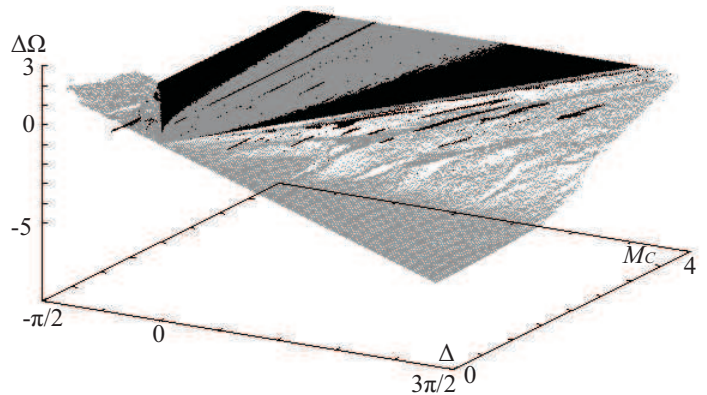}
\end{center}
\caption{Dependence of the deference of generators phases on the frequency detuning and coupling. Gray color responds non-regular regimes, black color responds periodic regimes; $\lambda=0$,
$\omega_{0}=\pi$, $\mu_{1}=\mu_{2}=0.9$.} \label{Fig.5}
\end{figure}

One more illustration of the phase synchronization in the system (\ref{2GQP}) is given in Fig.~\ref{Fig.5}. There is a three-dimensional surface, which gives the dependence of a difference of generators phases \footnote{In this case attractors is phase-coherent, and their phases can be determined geometrically \cite{r1}.} $\Delta\Omega$ on the frequency detuning and the coefficient of coupling. Moreover, non-regular regimes on this surface are marked by gray color; and periodic regimes are shown by black color. How we can see, the locking of quasiperiodic oscillations happens not only at weak coupling and frequency detuning, but at large values of this parameters.
\begin{figure}[h]
\begin{center}
\includegraphics[scale=0.5]{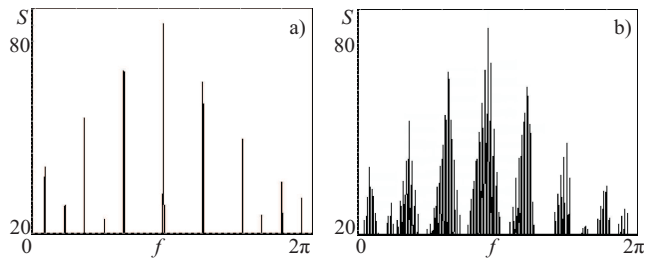}
\end{center}
\caption{Fourier spectrums of the system (\ref{2GQP}) at $\lambda=0$,
$\omega_{0}=\pi$, $\mu_{1}=\mu_{2}=0.9$, $M_{C}=0.07$; a)
$\Delta=0.05$ и b) $\Delta=0.1$.} \label{Fig.6}
\end{figure}

The tongues of complete synchronization on Fig.~\ref{Fig.2}~a) converge at the decrease of the coupling and has \emph{threshold} by coupling. It is mean, that when we go outside of the main two-frequency area at the small coupling, then will occur quasiperiodic regime of higher dimension, Fig.~\ref{Fig.2}~b). The features of this system consist of that at the small coupling from two-frequency torus birth four-frequency torus. The cause of this is that we consider generators identical by the second frequency parameter: $\mu_{1}=\mu_{2}$. In Fig.~\ref{Fig.6} the examples of the Fourier spectrums are shown, which illustrated such kind of the transition from two-frequency to four-frequency quasiperiodic oscillations. On Fig.~\ref{Fig.6}~a) it is clearly shown, that the spectrum has a character form for two-frequency quasiperiodicity: the main peak, which is surrounding by components-satellites. On Fig.~\ref{Fig.6}~b) is represented the spectrum for four-frequency quasiperiodic regime, where one can see that near each component of the spectrum for the two-frequency regime have birth new components-satellites.

Let us turn to Fig.~\ref{Fig.2}. On Fig~\ref{Fig.2}~b) one can see, that at the small coupling four-frequency tori dominate. On the other hand, there is  the set of tongues of three-frequency tori, the bottom of which is disposed on the axis $\Delta=0$. Tops of this tongues build up along line, which separates band of three-frequency regime occurring at the large coupling. In this band of three-frequency tori the set tongues of two-frequency quasiperiodicity is embedded. At the large coupling the regime of chaos and hyperchaos occur. In these areas of two-frequency quasiperiodicity, chaos and hyperchaos, there is the set of islands of periodic regimes of high order, but there is not any regular structures. All areas of periodic regimes have threshold by coupling.

\begin{figure}[h]
\begin{center}
\includegraphics[scale=0.5]{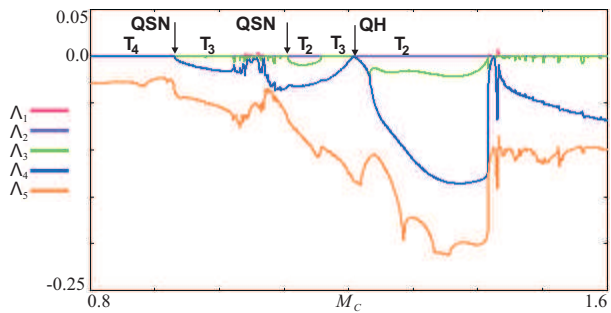}
\end{center}
\caption{Dependence of the largest Lyapunov exponentss on the coupling and the points of quasiperiodic bifurcations of different types at $\lambda=0$, $\omega_{0}=\pi$, $\mu_{1}=\mu_{2}=0.9$,
$\Delta=4.52$.} \label{Fig.7}
\end{figure}
Now let consider the Fig.~\ref{Fig.7}, in which a dependence of the largest Lyapunov exponents on the coupling for fixed frequency detuning $\Delta=4.52$ is shown. This dependence allows to describe the torus evolution with coupling increasing. How one can see, the areas of tori with different dimension alternate. For the occurrence of different many-dimensional tori respond some special bifurcations: quasiperiodic bifurcations \cite{r15}. In our system are possible two type of bifurcations. So, the saddle-node bifurcations of the torus happen in points \textbf{QSN}, moreover, this bifurcations can happen with two-frequency torus as like as three-frequency torus. In the point \textbf{QH} there is quasiperiodic bifurcation of Hopf-Neimark-Sacker. It is the bifurcation of the soft birth of three-frequency torus from two-frequency torus. To distinguish these two kind of bifurcations one can see by the dependence of Lyapunov exponents on the parameter. In the first case one of the exponent become zero, and equal zero after bifurcation, and the second exponent is always negative. For the Hopf quasiperiodic bifurcation the first and the second Lyapunov exponents before bifurcation have  the same value, and both become zero in the bifurcation point. Then the first Lyapunov exponent equals zero, and the second again goes to negative value \cite{r15}.

\section{Case of different frequencies parameters}
\begin{figure}[h]
\begin{center}
\includegraphics[scale=0.6]{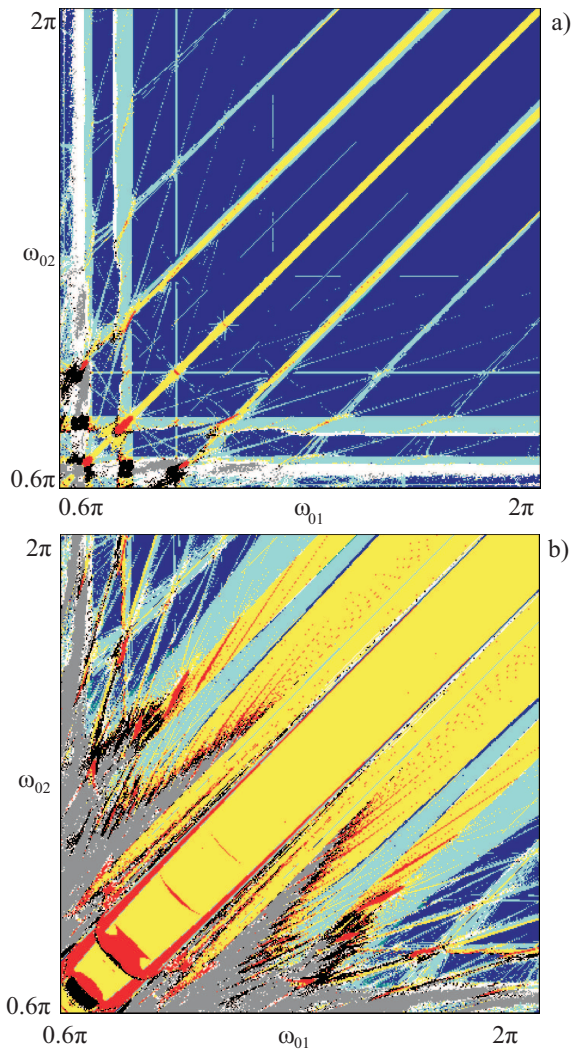}
\end{center}
\caption{Chart of Lyapunov exponents on the parameter plane of the base frequencies of generators; a) $M_{C}=0.05$, b) $M_{C}=0.5$; other parameters: $\lambda=0$, $\omega_{0}=\pi$, $\mu_{1}=\mu_{2}=0.9$.}
\label{Fig.8}
\end{figure}

How we note above, the system (\ref{2GQP}) has four frequency parameters, that is why we can consider the dynamics of the system on the another parameter planes. Let consider the dynamics on the parameter plane of base frequencies of generators ($\omega_{01}$, $\omega_{02}$). The parameters $\mu_{1}$ and $\mu_{2}$ fix equal and corresponding to the quasiperiodic regime $\mu_{1}=\mu_{2}=0.9$. On Fig.~\ref{Fig.8} the charts of the Lyapunov exponents at the large and weak coupling are shown. The color palette is the same as on Fig.~\ref{Fig.2}~b). How one can see at the large coupling and small frequencies in the system dominates hyperchaos. It is concerned with that autonomous subsystems demonstrate chaos, and the spectrum of Lyapunov exponents of each generator has one positive exponent. When we couple these generators, we get the system, which characterized by the two positive Lyapunov exponents.

On Fig.~\ref{Fig.8} is clearly shown the symmetry of the picture respect of the line of equal frequencies $\omega_{01}=\omega_{02}$, i.e. the oscillators is identical by the all another parameters. There is a band of two-frequency quasiperiodic regimes along diagonal $\omega_{01}=\omega_{02}$. This band corresponds to the area of quasiperiodic phase synchronization. It is interesting, that width of this band is constant if we will increase the frequencies $\omega_{01}$ and $\omega_{02}$.

At the weak coupling $M_{C}=0.05$  on Fig.~\ref{Fig.8}~b) the band of phase synchronization become very narrow. In this case on the chart clear shown the \emph{resonance Arnold web} \cite{r16}. It is the set of bands of two- and three-frequencies regimes embedded in the area of four-frequency regime. At the intersections of this bands the periodic regimes occur. Such structure of the parameter plane can de explained by the presence in the system different resonances.

\begin{figure}[h]
\begin{center}
\includegraphics[scale=0.6]{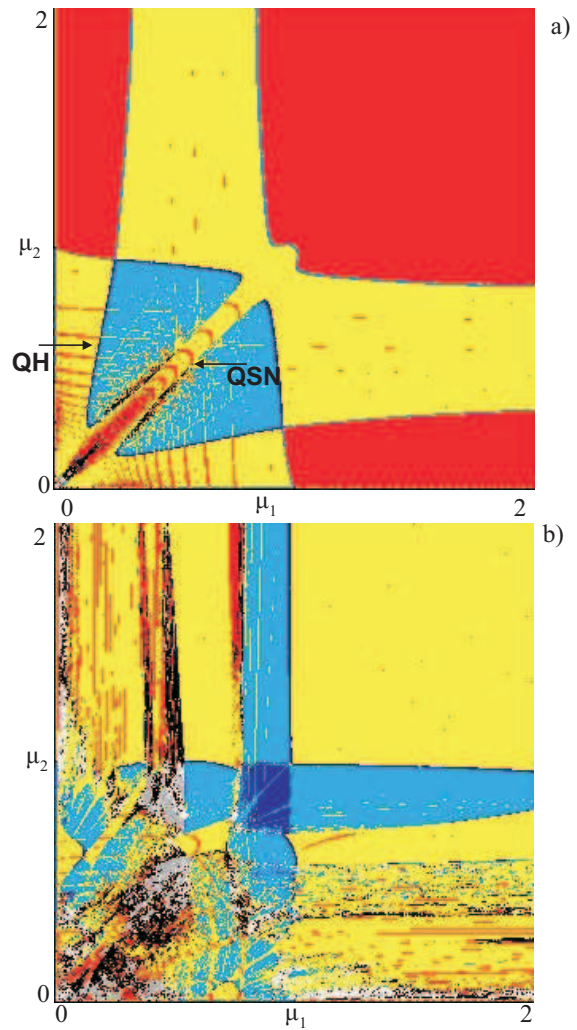}
\end{center}
\caption{Chart of Lyapunov exponents for system (\ref{2GQP});
$\lambda=0$, $\omega_{0}=\pi$, $\mu_{1}=\mu_{2}=0.9$, $M_{C}=0.1$,
a) $\Delta=0$, b) $\Delta=0.75$.} \label{Fig.9}
\end{figure}

On Fig.~\ref{Fig.9} are given the charts of Lyapunov exponents on the parameter plane of two another frequency parameters ($\mu_{1}$, $\mu_{2}$) at zero and non-zero frequency detuning. Along diagonal also one can see the area of two-frequency regimes in the form of band with embedded resonances of high order. This area of two-frequency tori is disposed inside island of three-frequency tori. One of the boundary of area of three-frequency quasiperiodicity corresponds quasiperiodic bifurcation Hopf-Neimark-Sacker \textbf{QH}, and the second by line of saddle-node bifurcation of two-frequency torus \textbf{QSN}. If we add some frequency detuning then near diagonal line occurs area of four-frequency tori.
\section{Conclusion}
The dynamics of two coupled generators of quasiperiodic oscillations gives the complex picture in the space of controlling parameters. There is can be phase synchronization of quasiperiodic oscillations as like as complete synchronization with locking of all frequency components. The complete synchronization has threshold by the coupling. In the picture of frequency detuning vs coupling there are the set of tongues of three-frequency tori, intersection of which form the band of three-frequency regimes. In this band are embedded areas of two-frequency quasiperiodic oscillations and islands of periodic regimes of high order. In the system can happen saddle-node bifurcations of different dimensional tori, and quasiperiodic Hopf-Neimark-Sacker bifurcations. On the parameter plane of base frequencies of generators occur the resonance Arnold web at the weak coupling. With increasing coupling Arnold web is destroying.

This research was supported, in part, by the grants of RFBR No.14-02-00085

\end{document}